\documentclass[showpacs,preprintnumbers,prb]{revtex4}
\usepackage{amsmath}
\usepackage{epic}
\usepackage{eepic}
\usepackage{graphicx}

\newcommand{\s}{{\bf S}}

\newcommand{\ket}[1]{\left|#1\right\rangle}
\newcommand{\kett}[1]{|#1\rangle}
\newcommand{\bra}[1]{\left\langle#1\right|}
\newcommand{\ketcol}[2]{\ket{\begin{subarray}{c}#1\\[4pt]#2\end{subarray}}}

\newcommand{\figref}[1]{Fig.~\ref{#1}}
\newcommand{\secref}[1]{Sec.~\ref{#1}}
\newcommand{\refl}{{\cal R }}

\begin{document}

\preprint{Phys. Rev. B \textbf{75}, 214421}

\title{Energy-level ordering and ground-state quantum numbers\\
for frustrated two-leg spin-1/2 ladder model}

\author{Tigran Hakobyan}
\email{hakob@yerphi.am}
\affiliation{%
Yerevan Physics Institute, 2 Alikhanyan Br., Yerevan, Armenia
\\
and
\\
Yerevan State University, 1 Alex Manoogian, Yerevan, Armenia
}%

\date{6 February 6 2007}

\begin{abstract}
The Lieb-Mattis  theorem about antiferromagnetic ordering of
energy levels on bipartite lattices is generalized to finite-size
two-leg spin-1/2 ladder model frustrated by diagonal interactions.
For reflection-symmetric model with site-dependent interactions we
prove exactly that the lowest energies in sectors with fixed total
spin and reflection quantum numbers are monotone increasing
functions of total spin. The nondegeneracy of most levels  is
proved also. We also establish the uniqueness and obtain the spin
value of the lowest-level multiplet in the whole sector formed by
reflection-symmetric (antisymmetric) states. For a wide range of
coupling constants, we prove that the ground state is a unique
spin singlet. For other values of couplings, it may be also a
unique spin triplet or may consist of both multiplets. Similar
results have been obtained for the ladder with arbitrary boundary
impurity spin. Some partial results have also been obtained in the
case of periodical boundary conditions.
\end{abstract}


\maketitle

\section{Introduction}

Nowadays, the frustrated spin systems are the subject of intensive
study. \cite{frus-book,Misg01,Moess01,Sch05} The interest on them
is stimulated by recent progress in synthesizing corresponding
magnetic materials. \cite{frus-exper} In these models, due to
competing interactions, the classical ground state cannot be
minimized locally and usually possesses a large degeneracy. The
frustration can be caused by the geometry of the spin lattice or
by the presence of both ferromagnetic and antiferromagnetic
interactions.

In contrast, the ground states of classical models on bipartite
lattices are N\'eel ordered and are unique up to the global spin
rotations. The bipartitness  means that the lattice can be divided
into two sublattices $A$ and $B$, so that all interactions within
the same sublattice are ferromagnetic while the interactions
between different sublattices are antiferromagnetic. For the
N\'eel state, all spins of $A$ have same direction while all spins
of $B$ are aligned in opposite direction.

The quantum fluctuations destroy N\'eel state and the quantum
model, in general, has more complicated ground state. However, for
bipartite spin systems, the quantum ground state inherits some
properties of its classical counterpart. In particular, Lieb and
Mattis proved that the quantum ground state of finite-size system
is a unique multiplet with total spin $S_{\text{gs}}=|S_A-S_B|$,
which coincides with the spin of the classical N\'eel state. Here,
$S_A$ and $S_B$ are the highest possible spins on corresponding
sublattices. \cite{LM62,Lieb89}  This feature of bipartite systems
looks natural, since in the limit of large spin values the quantum
model approaches to the classical one. Moreover, a simple general
rule that the energy increases with increasing spin, which is
conventional for classical antiferromagnets, makes a sense in this
case, too. More precisely, the lowest energy $E_S$ in the sector,
where the total spin is equal to $S$, is a monotone increasing
function of the spin for any $S\ge S_{\text{gs}}$. \cite{LM62}
This property is known as Lieb-Mattis theorem about
antiferromagnetic ordering of energy levels. Lieb-Mattis theorem
is very important because it provides information about the ground
state and spectrum of bipartite spin systems without exact
solution or numerical simulation. Recently, it has been
generalized to  SU($n$) symmetric quantum chain with defining
representation and nearest-neighbor interactions. \cite{H04} A
ferromagnetic ordering of energy levels has been formulated and
proven also for XXX Heisenberg chains of any spin. \cite{Nach03}

More limited number of exact results is available for frustrated
quantum spin systems. It is difficult to treat even their
classical ground states as was mentioned above. The usual
Lieb-Mattis theorem is no longer valid for frustrated systems.
Recently, however, Lieb and Schupp proved rigorously that a
reflection-symmetric spin system with antiferromagnetic crossing
bonds possesses at least one spin-singlet  ground state.
\cite{LS99,LS00} Moreover, under certain additional conditions,
all ground states become singlets. A lot of frustrated spin
systems satisfy the conditions of Lieb-Schupp approach.
Two-dimensional pyrochlore antiferromagnet having checkerboard
lattice structure is an example of such type of system.
\cite{LS99} Another example is well-known two-dimensional
antiferromagnetic $J_1-J_2$ model, namely, the Heisenberg model on
square lattice with diagonal interactions, which has been studied
recently. \cite{NT03,Bal04,Sind04,Bish98} Note that this approach
does not provide any information about the degeneracy of ground
states. Thus, for frustrated spin systems, this question still
remains open. We mention also that the method used in
Refs.~\onlinecite{LS99} and \onlinecite{LS00} is restricted to the
systems, which do not have a spin on their symmetry axis.

At the same time, recently, certain signs in favor of Lieb-Mattis
theorem for frustrated spin systems have appeared. For $J_1-J_2$
model, the number of arguments based on exact diagonalization and
spin-wave approximation had been presented in support of the fact
that the antiferromagnetic ordering of energy levels is preserved
under weak frustration caused by diagonal couplings. \cite{Rich95}
Recently, for the same model, the condition
$S_{\text{gs}}=|S_A-S_B|$ for the ground state spin has been
tested numerically in Ref.~\onlinecite{Liu02}. The authors came to
the conclusion that it remains true provided that the frustration
is sufficiently weak in order to destroy ground-state N\'eel
order. The Lieb-Mattis property had been observed also at finite
size spectrum of the Heisenberg model on the triangular and Kagome
lattices. \cite{Lech} All these investigations suggest that in
many cases, even a relatively strong frustration cannot destroy the
antiferromagnetic ordering of energy levels.

On the other hand, recently, the spin ladder systems have
attracted much attention. The interest to them has been generated
by significant progress made within the last years in fabrication
of such type of compounds having a structure similar to the
structure of two-dimensional high-temperature superconductors.
\cite{Dag99} Ladder systems are simpler for study than their more
complex two-dimensional counterparts. Moreover, the powerful
theoretical and numerical methods, elaborated for one-dimensional
models, can be applied to study them. The two-leg  ladder
frustrated by diagonal interactions is one of the simplest
frustrated spin systems. It can be viewed as a
quasi-one-dimensional analog of $J_1-J_2$ model. Together with
various generalizations, it has been investigated intensively
during the last decade.
\cite{Xian95,LT06,Kotov98,Wang00,Ners00,Kim00,Hakob01,frus-ladd}
The ground-state phase diagram consists of two topologically
distinct Haldane phases separated by the curve (or surface) of
phase transition.

Inspired by aforementioned activities, in this paper we generalize
the Lieb-Mattis ordering theorem to two-leg spin-1/2 ladder model
with diagonal interactions. We consider the system, which rests
invariant under the reflection with respect to the longitudinal
symmetry axis of the ladder. Up to our knowledge, this is the
first exact result related to the energy level ordering of a
frustrated spin system.

In \secref{sec:open}, we formulate and prove an analog of
Lieb-Mattis theorem for frustrated spin-1/2 ladder with free
boundaries.  The reflection symmetry spits the total space of
states into two invariant sectors, composed correspondingly from
reflection-symmetric and reflection-antisymmetric states. We
establish the antiferromagnetic ordering of energy levels in
symmetric and antisymmetric sectors separately for wide range of
the interaction constants. The nondegeneracy of most levels is
proved also. In \secref{sec:GS}, using the ordering rule and the
results of Lieb and Schupp \cite{LS99,LS00} on
reflection-symmetric spin systems, the total spin and reflection
quantum numbers of the ground state are derived. The results are
tested and compared with numerical simulations. In
\secref{sec:comparison}, the validity of obtained results is
checked in some particular cases, for which exact results are
already known. Section \ref{sec:impurity} is devoted to the
frustrated ladder with boundary impurity spin. Similar ordering of
energy levels is established in this case, too. Some partial
results are obtained for the model with periodic boundary
conditions in \secref{sec:periodic}. In the last section, we
briefly summarize the results obtained in this paper. In Appendix,
we apply the general results for reflection-symmetric spin systems
obtained in Refs.~\onlinecite{LS99} and \onlinecite{LS00} to the
frustrated ladder model.

\section{Energy level ordering of finite frustrated ladder with open boundaries}
\label{sec:open}
\subsection{Frustrated ladder model}
In this paper we consider two finite identical antiferromagnetic
spin-1/2 chains coupled by rung and diagonal interactions.
\cite{Kotov98,Wang00} The Hamiltonian reads (see
\figref{fig:ladder}):
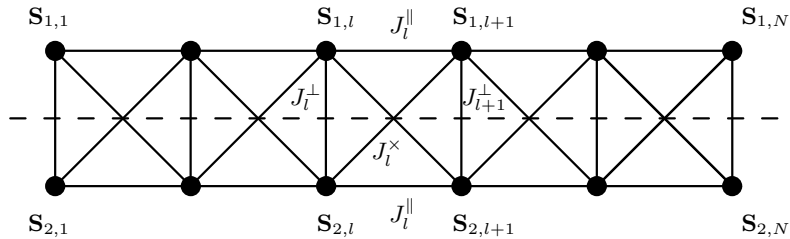
\begin{figure}[b]
\setlength{\unitlength}{1.2mm}
\begingroup\makeatletter\ifx\SetFigFont\undefined%
\gdef\SetFigFont#1#2#3#4#5{%
  \reset@font\fontsize{#1}{#2pt}%
  \fontfamily{#3}\fontseries{#4}\fontshape{#5}%
  \selectfont}%
\fi\endgroup%
{\renewcommand{\dashlinestretch}{30}

\begin{picture}(100,25)(-10,-5)
\allinethickness{1.000pt}%

\put(0,0){\blacken\ellipse{2}{2}}
\put(0,15){\blacken\ellipse{2}{2}}
\put(15,0){\blacken\ellipse{2}{2}}
\put(15,15){\blacken\ellipse{2}{2}}
\put(30,0){\blacken\ellipse{2}{2}}
\put(30,15){\blacken\ellipse{2}{2}}
\put(45,0){\blacken\ellipse{2}{2}}
\put(45,15){\blacken\ellipse{2}{2}}
\put(60,0){\blacken\ellipse{2}{2}}
\put(60,15){\blacken\ellipse{2}{2}}
\put(75,0){\blacken\ellipse{2}{2}}
\put(75,15){\blacken\ellipse{2}{2}}

\drawline(0,0)(75,0)
\drawline(0,15)(75,15)
\dashline{2.}(-5,7.5)(80,7.5)

\drawline(0,0)(0,15)
\drawline(15,0)(15,15)
\drawline(30,0)(30,15)
\drawline(45,0)(45,15)
\drawline(60,0)(60,15)
\drawline(75,0)(75,15)

\drawline(0,0)(15,15)
\drawline(0,15)(15,0)
\drawline(15,0)(30,15)
\drawline(15,15)(30,0)
\drawline(60,0)(75,15)
\drawline(60,15)(75,0)
\drawline(30,0)(45,15)
\drawline(30,15)(45,0)
\drawline(45,0)(60,15)
\drawline(45,15)(60,0)
\drawline(60,0)(75,15)
\drawline(60,15)(75,0)

%
%

\put(26,9){$J^\perp_l$}
\put(37,17){$J^\parallel_l$}
\put(37,-4){$J^\parallel_l$}
\put(35,3){$J^\times_l$}
\put(45,9){$J^\perp_{l+1}$}

\put(-3,18){$\s_{1,1}$}
\put(-3,-5){$\s_{2,1}$}
\put(29,18){$\s_{1,l}$}
\put(29,-5){$\s_{2,l}$}
\put(44,18){$\s_{1,l+1}$}
\put(44,-5){$\s_{2,l+1}$}
\put(76,18){$\s_{1,N}$}
\put(76,-5){$\s_{2,N}$}

\end{picture}
}
\caption{\label{fig:ladder} Frustrated ladder with site-dependent
couplings. Here $J^\parallel_l$  are intrachain couplings while
$J^\perp_l$ and  $J^\times_l$ are correspondingly rung and
diagonal interchain couplings. The dashed line is the symmetry
axis of the model.}
\end{figure}
\begin{equation}
\label{h}
\begin{split}
H=\sum_{l=1}^{N-1}
J^\parallel_l(\s_{1, l}\cdot\s_{1, l+1}+\s_{2, l}\cdot\s_{2, l+1})
+\sum_{l=1}^{N-1}
J^\times_l(\s_{1,l}\cdot\s_{2,l+1}+\s_{1,l+1}\cdot\s_{2,l})
+\sum_{l=1}^{N} J^\perp_l\s_{1,l}\cdot\s_{2,l},
\end{split}
\end{equation}
where $\s_{1,l}$ and $\s_{2,l}$ are the spin operators of the
first and second chains respectively.  $J^\perp_l$ and
$J^\times_l$ are rung and diagonal interchain couplings,  while
$J^\parallel_l$ are intrachain couplings. The open boundary
conditions are applied here. All couplings depend on site. Their
choice corresponds to a model possessing the reflection symmetry
$\refl$ with respect to the longitudinal symmetry axis. The
reflection just permutes the spins of two chains:
$\refl\,\s_{1,l}=\s_{2,l}\refl$.

We  do not put any restriction on the rung couplings $J_l^\perp$
and consider antiferromagnetic intrachain couplings only. We
suppose also that the diagonal spin interactions are weaker than
intrachain interactions, i.e.,
\begin{equation}
\label{range} J^\parallel_l>|J^\times_l|.
\end{equation}
Note that the condition above is not too restrictive for
antiferromagnetic values of $J^\times_l$ because the system
remains invariant under the exchange of rung and diagonal
couplings $J^\parallel_l \leftrightarrow J^\times_l$ belonging to
$l$th box. The resulting model corresponds  to a ladder obtained
by permutation $\s_{1,l'}\leftrightarrow\s_{2,l'}$ of two spins in
all rungs positioned on the right from $l$th rung (i.e., for all
$l'>l$). It is topologically equivalent to the original model.
Therefore, one can consider the couplings subjected to
$J^\parallel_l\ge J^\times_l$ only without loss of generality.
Note that the condition \eqref{range} excludes the values of
diagonal couplings $J^\times_l$ equal to $\pm J^\parallel_l$.

One must mention that we call the model \eqref{h} a
frustrated ladder, but, in fact, it becomes bipartite for some
couplings. In particular, it is bipartite for ferromagnetic
diagonal or rung interactions. These cases will be discussed in
\secref{sec:comparison}.

The Hamiltonian  preserves the total spin
$\s=\sum_{l}\s_{1,l}+\s_{2,l}$ of the system. Since $\refl$ and
$\s$ are compatible, the eigenstates of (\ref{h}) can be chosen to
be parametrized  by the spin ($S=0,1,\dots,N$), spin projection
($M=-N,-N+1,\dots,N$), and reflection ($\sigma=\pm1$) quantum
numbers.

In order to make easier the use of the reflection symmetry,
we introduce the symmetric and antisymmetric superpositions
of two spins on each rung:
\begin{equation}
\label{S-sa}
\s^{(s)}_l=\s_{1,l}+\s_{2,l}, \quad \s^{(a)}_l=\s_{1,l}-\s_{2,l}, \quad l=1,2,\dots,N.
\end{equation}
The symmetrized spin $\s^{(s)}_l$ describes the total spin of
$l$th rung. It remains unchanged under reflection, i.e.
$\refl\,\s^{(s)}_l\refl=\s^{(s)}_l$, while the antisymmetrized
rung spin acquires minus sign, i.e.,
$\refl\,\s^{(a)}_l\refl=-\s^{(a)}_l$.

Now we express the Hamiltonian (\ref{h}) in terms of these
operators. After omitting nonessential scalar term,
the Hamiltonian takes the following simple form:
\begin{equation}
\label{hsa} H=\sum_{l=1}^{N-1}(J^s_l\,
\s^{(s)}_l\cdot\s^{(s)}_{l+1}+J^a_l\,
\s^{(a)}_l\cdot\s^{(a)}_{l+1}) +\frac{1}{2}\sum_{l=1}^{N}
J^\perp_l(\s^{(s)}_l)^2.
\end{equation}
Here, we have introduced the symmetrized and antisymmetrized couplings, which are
antiferromagnetic due to the condition \eqref{range} imposed above on the intrachain
and diagonal couplings:
\begin{equation}
\label{Jsa}
J^s_l=\frac{J^\parallel_l+J^\times_l}{2}>0
\quad \text{and} \quad
J^a_l=\frac{J^\parallel_l-J^\times_l}{2}>0.
\end{equation}
Note that \eqref{hsa} does not contain terms, which mix symmetrized
and antisymmetrized spin operators. This fact is a consequence of
the reflection symmetry of the model. We mention also that similar
decomposition for frustrated ladder Hamiltonian was applied also
in Refs.~\onlinecite{Xian95} and \onlinecite{LT06}.

In terms of lowering and rising operators
$S^{(s)\pm}_l=S^{(s)x}_l\pm iS^{(s)y}_l$ and
$S^{(a)\pm}_l=S^{(a)x}_l\pm iS^{(a)y}_l $ the Hamiltonian reads
\begin{equation}
\label{h-spm}
\begin{split}
H&=\frac{1}{2}\sum_{l=1}^{N-1} (J^s_lS^{(s)+}_lS^{(s)-}_{l+1}+J^s_lS^{(s)-}_lS^{(s)+}_{l+1}
+J^a_lS^{(a)+}_lS^{(a)-}_{l+1}+J^a_lS^{(a)-}_lS^{(a)+}_{l+1})
\\
&+\sum_{l=1}^{N-1} (J^s_lS^{(s)z}_lS^{(s)z}_{l+1}+J^a_lS^{(a)z}_lS^{(a)z}_{l+1})
+\frac{1}{2}\sum_{l=1}^{N} J^\perp_l(\s^{(s)}_l)^2.
\end{split}
\end{equation}

Further, in this section, we will show that for
$J^\parallel_l>|J^\times_l|$ the minimal energy levels
$E_{S,\sigma}$ in the symmetric ($\sigma=1$) as well as in the
antisymmetric ($\sigma=-1$) spin-$S$ sectors are nondegenerate and
ordered antiferromagnetically, i.e., are increasing functions of
$S$. This is an extension of Lieb-Mattis ordering theorem
\cite{LM62} to the frustrated ladder model. Here, we briefly
outline the steps of the proof.

First, we construct a basis, in which all nonzero off-diagonal
matrix elements of the Hamiltonian become negative. In the next
step, we show that the matrix of Hamiltonian being restricted to
the subspace of states with fixed values of spin projection
$S^z=M$ and reflection $\refl=\sigma$ quantum numbers is
connected. Then according to Perron-Frobenius theorem, the
lowest-energy state in every such subspace, called a \emph{relative
ground state}, is nondegenerate. As we will show,  the relative
ground state in most cases has total spin value equal to absolute
value of its $z$ projection ($S=|M|$). Together with the total
spin symmetry of the Hamiltonian this implies that the multiplet,
to which the relative ground state belongs, possesses the lowest energy
value among all spin-$S$ multiplets. The antiferromagnetic
ordering between lowest energy levels $E_{S,\sigma}$ follows then
directly from their nondegeneracy.

\subsection{Nonpositive basis}

The existence of a basis, where all off-diagonal elements of the
Hamiltonian are nonpositive, is not obvious as it is for
nonfrustrated models. The spin flip applied on one sublattice
\cite{LM62,AL86} is adopted for bipartite systems and does not lead
to the desired basis for frustrated systems. In this subsection, we
construct nonpositive basis for the frustrated ladder model. The
first step is the use of the basis consisting of combined spin
states on each rung instead of the standard Ising basis consisting
of on-site spin-up and spin-down states. Then we apply a unitary
shift to the Hamiltonian, which makes all nonvanishing
off-diagonal elements arising from the symmetric part of
\eqref{hsa} negative. In the final step, all basic states are multiplied by
a sign factor making all nonvanishing off-diagonal matrix
elements negative.

It is rather difficult to trace out the sign of matrix elements of
$H$ in the Ising basis, consisting of on-site spin states. Instead,
we start with a more suitable basis consisting of combined spin
states on every rung. Any rung state can be expressed as a
superposition of three symmetric triplet states,
\begin{equation}
\label{triplet} \ket{1}:=\ket{1,1}=\ketcol{\uparrow}{\uparrow},
\qquad \ket{-1}:=\ket{1,-1}=\ketcol{\downarrow}{\downarrow}, \qquad
\ket{\tilde{0}}:= \ket{1,0}=\frac1{\sqrt{2}}
\left(\ketcol{\uparrow}{\downarrow}+\ketcol{\downarrow}{\uparrow}\right),
\end{equation}
and one antisymmetric singlet state,
\begin{equation}
\label{singlet}
\ket{0}=\frac1{\sqrt{2}}\left(\ketcol{\uparrow}{\downarrow}-\ketcol{\downarrow}{\uparrow}\right).
\end{equation}
We have used above standard notations $\uparrow,\downarrow$ for
on-site spin-up and spin-down states. For convenience, we mark the
triplet states shortly by labels $\pm1$ and $\tilde0$.

The total space of states is spanned by the basic states
\begin{equation}
\label{basis}
\ket{m_1}\otimes\ket{m_2}\otimes\ldots\otimes\ket{m_{N}}, \quad
 m_l=\pm1, \tilde0,0.
\end{equation}
Here $\ket{m}$ is one of the four rung states defined above. The
reflection operator $\refl$ is diagonal in this basis. Its quantum
number is $(-1)^{N_0}$, where $N_0$ is the number of singlets in
\eqref{basis}.

Now define unitary operator, which rotates the odd-rung spins
around the $z$ axis on angle $\pi$, as follows:
\begin{equation}
\label{U} U=\exp\left(i\pi
\sum_{l=1}^{[(N+1)/2]}S^{(s)z}_{2l-1}\right),
\end{equation}
where by $[x]$ we have denoted the integer part of $x$. Under the
action of $U$, the odd-rung lowering-rising operators change the
sign ($US^{(s,a)\pm}_{2l-1}U^{-1}=-S^{(s,a)\pm}_{2l-1}$), while the
others remain unchanged. Recall that for bipartite models, a
similar unitary shift applied to the spins of one sublattice makes
all off-diagonal elements of the Hamiltonian nonpositive.
\cite{LM62,AL86} In our case, the Hamiltonian (\ref{h-spm})
transforms to
\begin{equation}
\label{h-U}
\begin{split}
\tilde H = UHU^{-1} &=\frac{1}{2}
\sum_{l=1}^{N-1} (-J^s_lS^{(s)+}_lS^{(s)-}_{l+1}-J^s_lS^{(s)-}_lS^{(s)+}_{l+1}
-J^a_lS^{(a)+}_lS^{(a)-}_{l+1}-J^a_lS^{(a)-}_lS^{(a)+}_{l+1})
\\
&+\sum_{l=1}^{N-1} (J^s_lS^{(s)z}_lS^{(s)z}_{l+1}+J^a_lS^{(a)z}_lS^{(a)z}_{l+1})
+\frac{1}{2}\sum_{l=1}^{N} J^\perp_l(\s^{(s)}_l)^2.
\end{split}
\end{equation}
It is easy to see that $J^\perp$ part of the Hamiltonian,
presented by the last sum in (\ref{h-U}), is diagonal in the basis
\eqref{basis} because it consists of the squares of rung spin
operators.

The next observation is that $J^{(s)}$ part of the Hamiltonian
$\tilde H$ gives rise to negative off-diagonal elements. Indeed,
the symmetrized spin operators  describe the spin of rung state
and act separately on singlet and triplet states in the usual way.
All their off-diagonal matrix elements are positive:
\begin{equation}
\label{ssym}
\bra{\tilde{0}}S^{(s)+}\ket{-1}=\bra{1}S^{(s)+}\ket{\tilde{0}}
=\bra{-1}S^{(s)-}\ket{\tilde{0}}=\bra{\tilde{0}}S^{(s)-}\ket{1}
=\sqrt{2}.
\end{equation}
Taking into account the fact that the coefficients $J^s_l$ are
antiferromagnetic [see \eqref{Jsa}], it is easy to see that all
nonvanishing off-diagonal elements of the shifted Hamiltonian
\eqref{h-U}, which are generated by terms containing symmetrized
spin operators, are negative.

Finally, we consider the matrix elements produced by the
antisymmetric local terms of Hamiltonian \eqref{h-U}. In contrary
to symmetric case, the antisymmetrized spin operators mix triplet
and singlet states. All their nonzero matrix elements are
off-diagonal and have the following values: \cite{Noak06}
\begin{equation}
\begin{aligned}
\label{sasym}
\bra{0}S^{(a)+}\ket{-1}=\bra{-1}S^{(a)-}\ket{0}=\sqrt{2},
\quad
\bra{1}S^{(a)+}\ket{0}=\bra{0}S^{(a)-}\ket{1}=-\sqrt{2},
\quad
\bra{\tilde0}S^{(a)z}\ket{0}=\bra{0}S^{(a)z}\ket{\tilde 0}=1.
\end{aligned}
\end{equation}
Using the equations above, we can obtain all nontrivial matrix
elements generated by $J^{(a)}$ part of the Hamiltonian. The
nontrivial action of terms with operators $S^{(a)\pm}$ on two
adjacent spins is
\begin{subequations}
\label{spm}
\begin{align}
\label{spm-a}
& S^{(a)\mp}_1 S^{(a)\pm}_2\ket{\pm1}\otimes\ket{\mp1}=-2\;\ket{0}\otimes\ket{0},
&& S^{(a)\pm}_1 S^{(a)\mp}_2\ket{0}\otimes\ket{0}=-2\;\ket{\pm1}\otimes\ket{\mp1},
\\
\label{spm-b}
& S^{(a)\pm}_1 S^{(a)\mp}_2\ket{0}\otimes\ket{\pm1}=2\;\ket{\pm1}\otimes\ket{0},
&& S^{(a)\mp}_1 S^{(a)\pm}_2\ket{\pm1}\otimes\ket{0}=2\;\ket{0}\otimes\ket{\pm1}.
\end{align}
\end{subequations}
The action of terms in \eqref{h-U} containing $S^{(a)z}$ reads
\begin{subequations}
\label{sz}
\begin{align}
\label{sz-a}
& S^{(a)z}_1 S^{(a)z}_2\ket{0}\otimes\ket{0}=\ket{\tilde{0}}\otimes\ket{\tilde{0}},
& S^{(a)z}_1 S^{(a)z}_2\ket{\tilde{0}}\otimes\ket{\tilde{0}}=\ket{0}\otimes\ket{0},
\\
\label{sz-b}
& S^{(a)z}_1 S^{(a)z}_2\ket{\tilde{0}}\otimes\ket{0}=\ket{0}\otimes\ket{\tilde{0}},
& S^{(a)z}_1 S^{(a)z}_2\ket{0}\otimes\ket{\tilde{0}}=\ket{\tilde{0}}\otimes\ket{0}.
\end{align}
\end{subequations}
The subscripts 1 and 2 indicate the rung, on which the spin operator
acts.

Taking into account the positivity of antisymmetrized couplings
$J^{(a)}_l$ [see \eqref{Jsa}], we conclude that the elements
generated by \eqref{spm} acquire an overall minus sign in the
shifted Hamiltonian  matrix \eqref{h-U}. At the same time, the
elements corresponding to \eqref{sz} enter in the Hamiltonian with
the same sign. Hence, only the contribution from \eqref{spm-b}
gives rise to negative matrix elements in the Hamiltonian $\tilde H$,
whereas \eqref{spm-a} and \eqref{sz} are responsible for
unwanted positive off-diagonal elements. In order to alter their
sign, we make the following observation.

Due to the reflection symmetry, the parity of singlet states, i.e.,
$(-1)^{N_0}$, is conserved quantity under the action of each term
in \eqref{h-U}. The matrix elements \eqref{spm-a} and \eqref{sz-a}
are the only ones, which are responsible for the creation and
annihilation of a singlet pair. Other elements rest singlet
number unchanged. Thus, multiplying the basic states \eqref{basis}
on sign factor $(-1)^{\text{number of singlet
pairs}}=(-1)^{[N_0/2]}$, one can make the elements arising from
\eqref{spm-a} and \eqref{sz-a} negative. The sign factor does not
affect on other matrix elements.

The remaining action \eqref{sz-b} just permutes the singlet and
$S^z=0$ triplet states. In order to make this term negative,
introduce the ordering between pair of rung states $\ket{0}$ and
$\ket{\tilde0}$ inside multirung state \eqref{basis}. We say that
a pair is ordered if $\ket{\tilde0}$ is located on the left-hand
side from $\ket{0}$. Denote by  $N_{0\tilde0}$ the number of
disordered pairs in \eqref{basis}, i.e., the number of pairs
$(\ket{0},\ket{\tilde0})$, where $\ket{0}$ is on the left-hand
side from $\ket{\tilde 0}$. Note that the nearest-neighbor actions
\eqref{sz-b} change the ordering of only one pair. Therefore, if
we multiply a basic state by another sign factor
$(-1)^{N_{0\tilde0}}$, all matrix elements generated by
\eqref{sz-b} will change the sign and become negative. At the same
time, other matrix elements will hold unchained. Indeed, it is
easy to see, that the nearest-neighbor permutations
$\ket{\tilde0}\ket{\pm1}\leftrightarrow\ket{\pm1}\ket{\tilde0}$
and $\ket{0}\ket{\pm1}\leftrightarrow\ket{\pm1}\ket{0}$ do not
change the number of disordered pairs $N_{0\tilde0}$ while pair
creations and annihilations
($\ket{\pm1}\ket{\mp1}\leftrightarrow\ket{\tilde0}\ket{\tilde0}$
and $\ket{\pm1}\ket{\mp1}\leftrightarrow\ket{0}\ket{0}$) change
its value on \emph{even} number. Note that similar type of sign
factor has been used in order to prove uniqueness of relative
ground states for Heisenberg chains with higher symmetries.
\cite{Li01,H04}

Finally, the basis, in which all off-diagonal matrix elements of
the Hamiltonian \eqref{h-U} are nonpositive, is
\begin{equation}
\label{new-basis}
\ket{m_1,m_2,\ldots,m_{N}}
:=(-1)^{[N_0/2]+N_{0\tilde0}}
\ket{m_1}\otimes\ket{m_2}\otimes\ldots\otimes\ket{m_{N}}.
\end{equation}

\subsection{Relative ground states in $S^z=M$, $\refl=\sigma$ subspaces}

Due to the spin projection and reflection symmetries, the
Hamiltonian is invariant on each subspace with the definite values
of spin projection and reflection operators: $S^z=M$,
$\refl=\sigma$, where $M=-N,-N+1,\dots,N$ and $\sigma=\pm1$.
Keeping the terminology, we call it $(M,\sigma)$ subspace. Below,
we outline the proof that the matrix of the Hamiltonian in  basis
\eqref{new-basis} being restricted on every $(M,\sigma)$ subspace
is connected.

Note that due to \eqref{Jsa} all local actions considered in
\eqref{spm}, \eqref{sz} contribute in the Hamiltonian
\eqref{h-spm} with nonvanishing coefficients. It is easy to verify
using \eqref{ssym}, \eqref{spm}, and \eqref{sz} that any two
adjacent states $\ket{m_l}\otimes\ket{m_{l+1}}$ and
$\ket{m'_{l}}\otimes\ket{m'_{l+1}}$ are connected by the $l$th
local terms of the Hamiltonian \eqref{h-U} provided that their
quantum numbers are subjected to the conservation laws. In other
words, both states must possess the same spin projection and
reflection quantum numbers. This rule can be generalized by
induction to any two basic states  \eqref{h-U}. In fact, any
symmetric ($\sigma=1$) basic state \eqref{new-basis} after
successive applications of the local terms in \eqref{h-U} can be
transformed to state
$
|\underbrace{\pm1,\ldots,\pm1}_{|M|}\,,{\tilde0,\ldots},\tilde0\rangle
$,
where plus (minus) sign holds for positive (negative) values of
$M$. Similarly, any antisymmetric ($\sigma=-1$) state is connected
to state $
|\underbrace{\pm1,\ldots,\pm1}_{|M|}\,,{\tilde0,\ldots,\tilde0},0\rangle
$. This finishes the proof of the connectivity.

Now all conditions of Perron-Frobenius theorem \cite{PF} are
fulfilled and one comes to the following result. \emph{
\begin{itemize}
\item
The relative ground state $\ket{\Omega}_{M,\sigma}$ of $\tilde H$ in ($M,\sigma$) subspace
is unique and is a positive superposition of all basic states:
\begin{equation}
\label{gs} \ket{\Omega}_{M,\sigma}=\sum_{\substack{\sum_lm_l=M\\
(-1)^{N_0}=\sigma}} \omega_{m_1\dots
m_N}\ket{m_1,m_2,\ldots,m_{N}}, \qquad \omega_{m_1\dots m_N}>0.
\end{equation}
\end{itemize}
}

The state $\ket{\Omega}_{M,\sigma}$ must have a definite value
$S_{M,\sigma}$ of total spin quantum number. Otherwise, it could be
presented as a superposition of independent states with different
spins but  the same  energy. This would be in contradiction with
the uniqueness condition established above. It is evident that if
some spin-$S$ state overlaps with the relative ground state
$\ket{\Omega}_{M,\sigma}$
then $S_{M,\sigma}=S$.
Below, we use this property in order to determine $S_{M,\sigma}$.

Due to the spin reflection symmetry $S_{M,\sigma}=S_{-M,\sigma}$.
So, one can consider only non-negative values of $M$. First, we
suppose that $M>0$.

If  $\sigma=(-1)^{N-M}$, then the state
$\ket{\phi}=\kett{\underbrace{1,\ldots,1}_M,\underbrace{0,\ldots,0}_{N-M}}$,
which is the highest weight state for a spin $S=M$ multiplet,
contributes in the sum \eqref{gs}. Therefore, it overlaps with the
relative ground state. According to the arguments above,
$S_{M,\sigma}=M$ for this case.

 Else if $\sigma=(-1)^{N-M-1}$, then  both states
$\kett{\tilde0,\underbrace{1,\dots,1}_M,\underbrace{0,\dots,0}_{N-M-1}}$
and
$\kett{1,\tilde0,\underbrace{1,\dots,1}_{M-1},\underbrace{0,\dots,0}_{N-M-1}}$
are also presented in the decomposition of
$\ket{\Omega}_{M,\sigma}$ with \emph{positive} coefficients.
According to  the definition \eqref{new-basis} of basic states,
their sum, up to a nonessential sign factor, can be presented as
$\ket{\psi}=\sqrt{1/2}\left(\ket{1}\otimes\ket{\tilde0}+
\ket{\tilde0}\otimes\ket{1}\right)\otimes
\ket{1,\dots,1,0,\dots,0}$ and, of course, overlaps with
\eqref{gs}. In fact, $\ket{\psi}$ has a definite spin value, which
we will determine now. Remember we are currently working with the
Hamiltonian \eqref{h-U}, obtained from the original one by unitary
shift  \eqref{U}. Turning back to $H$ representation, we have to
shift the states back. Thus, the original state is
$U^{-1}\ket{\psi}=U\ket{\psi}\sim
\sqrt{1/2}\left(-\ket{1}\otimes\ket{\tilde0}+\ket{\tilde0}\otimes\ket{1}\right)\otimes
\kett{\underbrace{1,\dots,1}_{M-1},0,\dots,0}$
up to an overall sign factor. The first term of the product is the highest state of
the triplet in the decomposition of two triplets, while the second
term is the highest state of $S=M-1$ multiplet. Together they form
an $S=M$ multiplet. Therefore, $S_{M,\sigma}=M$ in the case of
$\sigma=(-1)^{N-M-1}$, too.

Consider now the two subspaces with $M=0$. If $\sigma=(-1)^{N}$,
the state constructed from rung singlets participates in the sum
\eqref{gs}, so the relative ground state is a spin singlet too:
$S_{0,\sigma}=0$. If $\sigma=(-1)^{N-1}$, the triplet state
$\kett{\tilde0,\underbrace{0,\dots,0}_{N-1}}$ enters in the sum,
and, therefore, the relative ground state is a triplet, i.e.,
$S_{0,\sigma}=1$.

Summarizing the results, we arrive at the following conclusion:
 \emph{
\begin{itemize}
\item
The relative ground state $\ket{\Omega}_{M,\sigma}$ always belongs to a spin-$|M|$ multiplet
except  $M=0$ and $\sigma=(-1)^{N-1}$ case when it belongs to a spin-triplet. In other words,
\begin{equation}
\label{Srelgs}
S_{M,\sigma}=\begin{cases}
1 & \text{if $M=0$ and $\sigma=(-1)^{N-1}$}\\
|M| & \text{for other values of $M$ and $\sigma$.}
\end{cases}
\end{equation}
\end{itemize}
}

\subsection{Ordering rule among the lowest levels in sectors with different value
of total spin}

Consider now the restriction of the Hamiltonian to the sector
containing states with fixed values of the total spin and reflection
quantum numbers.
Apparently, the relative ground state $\ket{\Omega}_{M,\sigma}$ is also
the lowest-energy state in
the sector characterized by $S=S_{M,\sigma}$ and $\refl=\sigma$.
Denote by $E_{S,\sigma}$  the lowest energy level in that sector.
According to the previous subsection, the relative ground state is unique on $(M,\sigma)$ subspace.
Therefore, the level $E_{S,\sigma}$ with $S=S_{M,\sigma}$ must be nondegenerate, i.e.,
it must contain only one
spin-$S$ multiplet with parity $\sigma$. This is just the multiplet, which encloses
the state $\ket{\Omega}_{M,\sigma}$ itself. Using \eqref{Srelgs}, it is easy to see
that any level  $E_{S,\sigma}$ is nondegenerate in this regard, except perhaps
the one with $S=0$ and $\sigma=(-1)^{N-1}$.

Moreover, every $(M,\sigma)$ subspace contains a representative
from any  multiplet with  parity $\sigma$ and spin  $S\ge |M|$.
Therefore, the state $\ket{\Omega}_{M,\sigma}$ has the minimum energy
for all these multiplets. According to \eqref{Srelgs},
its spin has the value $S_{M,\sigma}=|M|$
provided that $\sigma=(-1)^N$. The uniqueness
of relative ground state then implies that all levels $E_{S,\sigma}$ with spin $S>S_{M,\sigma}$
are higher than the level with  $S=S_{M,\sigma}$. Consequently,
$E_{S,\sigma}$ is a monotone increasing function of $S$.
If $\sigma=(-1)^{N-1}$, then the last equation is true for $|M|> 0$
only and, hence, $E_{S,\sigma}$ increases in the range $S\ge1$.

Finally, we arrive at the following conclusion.
\emph{
\begin{itemize}
\item
The minimum energy levels $E_{S,\sigma}$ are nondegenerate (except perhaps the one with
$S=0$ and $\sigma=(-1)^{N-1}$) and are ordered according to the  rule:
\begin{equation}
\label{ordering}
E_{S_1,\sigma}>E_{S_2,\sigma}:
\quad
\begin{cases}
\text{for $\sigma=(-1)^N$}       & \text{if $S_1>S_2$  }\\
\text{for $\sigma=(-1)^{N-1}$}   & \text{if either $S_1>S_2\ge 1$  or  $S_1=0$, $S_2=1$.}
\end{cases}
\end{equation}
\end{itemize}
}
The ordering rule above enables to determine the total spin of the minimum-energy
states in the symmetric and antisymmetric sectors.
\emph{
\begin{itemize}
\item
The ground state in entire $\sigma=(-1)^{N}$ sector is a spin singlet while in $\sigma=(-1)^{N-1}$
sector is a spin triplet. In both cases, it is unique.
\end{itemize}
}
In other words, for frustrated ladder with even number of rungs, the ground state in the symmetric
sector is a singlet state, while in the antisymmetric sector, it is formed by the three states
of a triplet. In contrary, for odd number of rungs, it is a singlet in the antisymmetric
sector, while in the symmetric sector, it consists of three triplet states.

\section{Ground state}
\label{sec:GS}

\subsection{Exact results}
The ordering rule \eqref{ordering}  does not compare energy levels
in symmetric ($\sigma=1$) sector with the levels in antisymmetric
($\sigma=-1$) sector. It is clear that the total ground state
coincides with the minimum-energy state in either symmetric or
antisymmetric sector. If the lowest levels in both sectors
coincide, the total ground state becomes degenerate. Using also the
results of previous section, we come to the following  conclusion.
\emph{
\begin{itemize}
\item The ground state of frustrated ladder with couplings obeying
\eqref{range} may be: \,\emph{(a)} a unique $\sigma=(-1)^{N}$
spin singlet; \,\emph{(b)} a unique $\sigma=(-1)^{N-1}$
spin triplet; \,\emph{(c)} any superposition of both of them, i.e.,
singlet\,$\oplus$\,triplet.
\end{itemize}
}
The frustrated ladder \eqref{h} in certain parameter region
belongs to a large class of reflection-symmetric models considered
recently by Lieb and Schupp. \cite{LS00} In Appendix, the rigorous
results obtained in Ref.~\onlinecite{LS00} are applied to the
frustrated spin ladder case. It appears that \emph{all} ground
states of the Hamiltonian \eqref{h} are spin \emph{singlets} if
the couplings satisfy $ J^\perp_l> |J^\times_{l-1}|+|J^\times_l|$,
where $J^\times_0=J^\times_N=0$ is supposed. The nonstrict
inequality sign can be set instead of strict one for
antiferromagnetic values of diagonal couplings. The details are
given in Appendix. Note that the aforementioned result is true for
any values of intrachain couplings $J^\parallel_l$ including
ferromagnetic ones.

From the other side, for $J^\parallel_l>|J^\times_l|$ our results
established in previous section become valid too. Setting together
both Lieb-Schupp and our results, we come to the conclusion that
\emph{only} case (a) above can take place. This means that the
ground state is a unique $\sigma=(-1)^N$ singlet, provided that
the couplings satisfy
$$
J^\perp_l> |J^\times_{l-1}|+|J^\times_l|
\quad
\text{and}
\quad
J^\parallel_l>|J^\times_l|.
$$

Consider separately the most familiar case of ladder with
site-independent couplings. According to the discussions above and
in Appendix, we come to the following conclusion.
\emph{
\begin{itemize}
\item The ground state of finite size frustrated ladder model with
antiferromagnetic rung $J^\perp$ and intrachain $J^\parallel$
interactions is a unique spin singlet with $\sigma=(-1)^N$
reflection symmetry if the couplings satisfy $J^\perp\ge
2J^\times>-J^\perp$ and $J^\parallel> |J^\times|$.
\end{itemize}
} It must be emphasized that this result is rigorous for finite
size ladders only, which are the main objects of study in this
paper.

\subsection{Thermodynamic limit, comparison with other approaches and numerical test}

\begin{figure}[t]
\begin{minipage}[t]{0.48\textwidth}
\includegraphics[width=\textwidth]{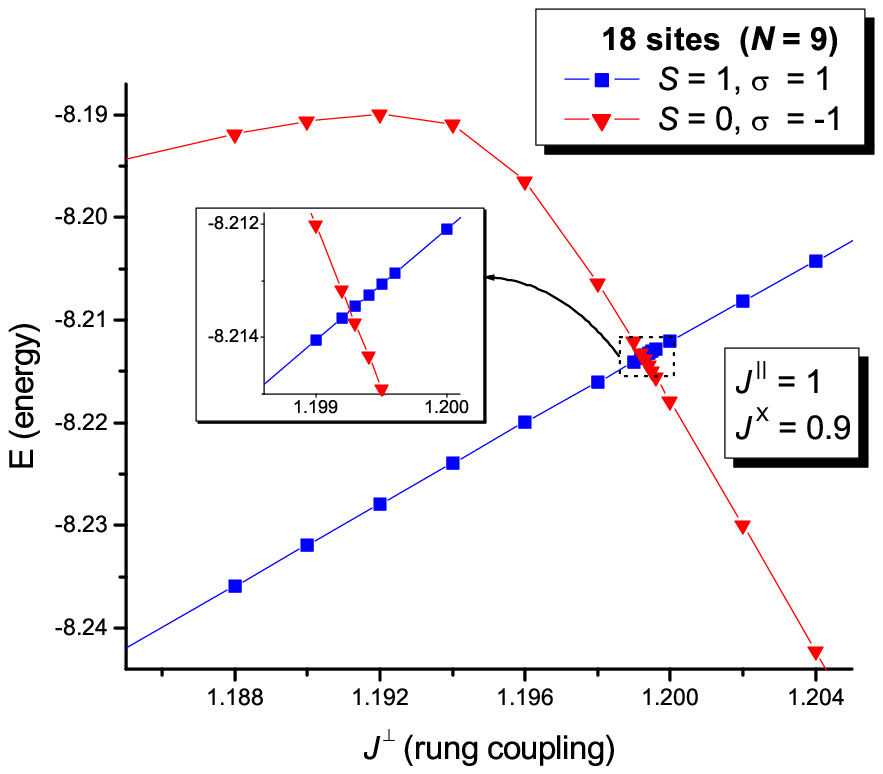}
\caption{\label{fig:crossing18} Two lowest energy levels, obtained
by exact diagonalization of frustrated ladder with 18 sites
($N=9$), are plotted as a function of rung coupling. Corresponding
states have different spin $S$ and reflection $\sigma$ quantum
numbers. The level crossing happens at some point.}
\end{minipage}
\hspace{0.02\textwidth}
\begin{minipage}[t]{0.48\textwidth}
\includegraphics[width=\textwidth]{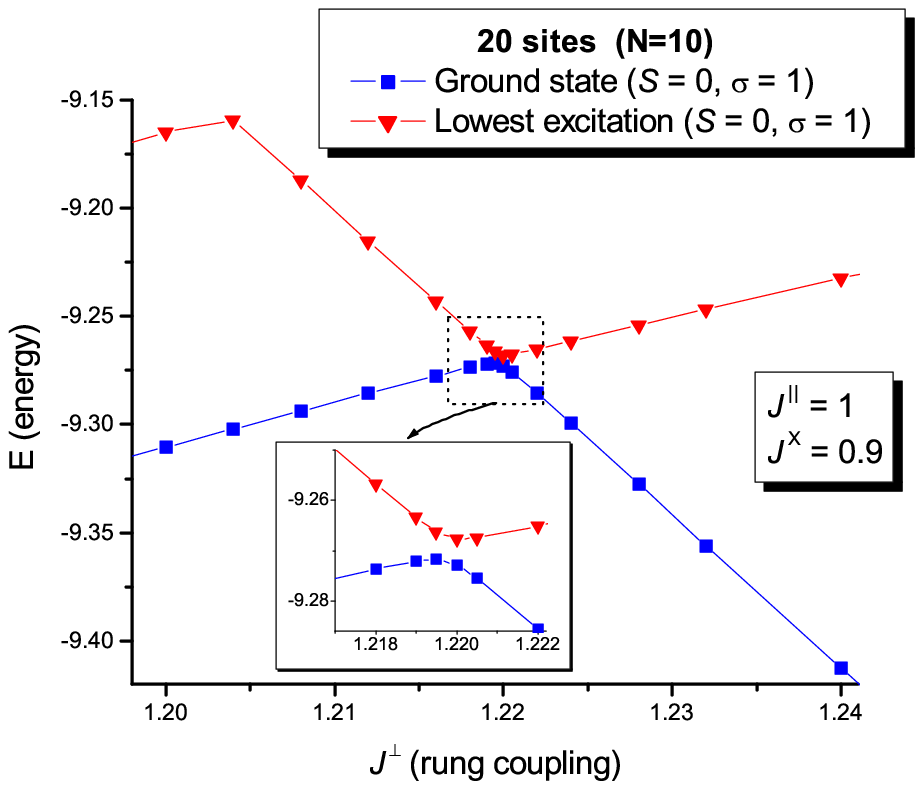}
\caption{\label{fig:crossing20} The ground state and lowest
excitation of frustrated ladder with 20 sites ($N=10$) obtained by
exact diagonalization for different values of rung coupling. Both
states have the same quantum numbers. The level crossing does not
happen in agreement with our exact results.}
\end{minipage}
\end{figure}

The results obtained above are in a good agreement with the ground-state
properties of frustrated ladder model, which have been
investigated intensively in the literature so far. From our
results, it is unclear whether or not the singlet-triplet
degeneracy takes place for finite-size ladders. Moreover, in the
thermodynamic limit $N\to\infty$, the additional degeneracy of
energy levels can occur and, in principle, the strict inequality
in \eqref{ordering} must be replaced by nonstrict one. The
investigations carried out by different methods suggest that all
three possibilities for the ground state described in the previous
subsection may take place.
In fact, the ground-state degeneracy happens at critical
points only, while in most cases, the ground state is a unique
$\sigma=(-1)^N$ singlet or $\sigma=(-1)^{N-1}$ triplet. This
property remains true in the thermodynamic limit. The rest of this
section is devoted to the testing of our exact results and more
detailed comparison with data obtained by other methods.

The weak interchain coupling analysis ($J^\perp,J^\times\ll
J^\parallel$) based on conformal field theory approach shows that
for $J^\perp>2J^\times$, the ground state has a tendency to form
singlets along the rungs and triplets along the diagonals, while
for $J^\perp<2J^\times$, it has a tendency to form triplets along
the rungs and singlets along the diagonals. \cite{Ners00} The
former ground state corresponds to the usual ladder or rung-dimer
phase, while the last one corresponds to spin-$1$ Haldane chain. In
both cases, it is nondegenerate. The two phases belong to the same
universality class because a continuous path connecting them
exists and it does not  contain any critical point. \cite{White96}
The distinction between them has a topological character.
\cite{Kim00,Hakob01} For $J^\perp=2J^\times$, both types of
dimerized ground states exist leading to twofold degeneracy.
\cite{Ners00}

 Numerical simulations \cite{Wang00,Hakob01}
strongly support this picture and extend the phase transition
curve out of the weak coupling region. The transition curve obeys
the relation $J^\perp<2J^\times$ and approaches to the line
$J^\perp=2J^\times$ at the weak coupling limit.

The properties of the ground state differ for ladders with odd and
even values of the rung number. First, we consider odd-rung
ladders. Small values of rung coupling $J^\perp$ give rise  to
Haldane-type ground state, which is a spin triplet with $\sigma=1$
[case (b) in the previous subsection] while large values lead to
rung-singlet-type ground state with $\sigma=-1$ [case (a)]. At
some intermediate point, the level crossing happens, which
corresponds to aforementioned case (c) with degenerate
singlet-triplet ground state. The level crossing point approaches
to the critical line in the thermodynamic limit. The numerical
investigation of small size ladder Hamiltonians supports this
picture. In \figref{fig:crossing18}, the two lowest levels of
frustrated ladder with 18 sites ($N=9$) obtained by exact
diagonalization are plotted for different values of rung coupling.
The figure clearly indicates that both levels cross at some point.

On the contrary, for even-rung ladders, both Haldane and rung-dimmer
states are singlets and located in $\sigma=1$ sector. The level
crossing is forbidden in this case because our results exclude the
existence of degenerate double-singlet ground state for finite-size systems.
In \figref{fig:crossing20}, we present the two lowest
levels obtained by numerical diagonalization of the system with 20
sites ($N=10$) close to the composite spin point
$J^\parallel=J^\times$. Both levels approach each other but do not
intersect. The level crossing is absent. It occurs at the
composite spin point, which is out of the parameter space
\eqref{range} considered in this paper. This model will be
discussed in the next section. In the thermodynamic limit, the
closest point of both levels approaches to the level crossing
point of odd-rung ladders and the level crossing picture recovers.

Similar behavior is observed for ferromagnetic interchain
couplings ($J^\perp<0$ and $J^\times<0$). Large values of
$|J^\perp|$ correspond to Haldane phase and its small values put
the system in rung-dimmer phase. The phase transition curve in the
thermodynamic limit is remarkably close to the line
$J^\perp=2J^\times$ (Ref. \onlinecite{ZWCh00})
and coincides with it in the
weak interchain interaction limit. \cite{Ners00} This curve has
been traced out with high precision by applying different boundary
conditions for each phase. \cite{Hakob01} A slight deviation from
it is observed close to the ferromagnetic phase only.

\section{Comparison with known results}
\label{sec:comparison}
In this section, we compare the properties of the generalized
ladder Hamiltonian \eqref{h}, \eqref{range} established in the
previous section with those obtained earlier for some particular
values of couplings.
\subsection{Ferromagnetic rung interactions}
If we set all rung couplings to be ferromagnetic and
others to be antiferromagnetic, the ladder model \eqref{h}
becomes \emph{bipartite}. The two sublattices $A$ and
$B$, forming the ladder, consist of spins of even and odd rungs
correspondingly. Then the conditions of standard Lieb-Mattis
ordering theorem \cite{LM62} are fulfilled. The Ising basis
after the unitary shift \eqref{U} becomes negative. Thus the
relative ground state $\ket{\Omega}_{M}$ is unique in whole
subspace $S^z=M$. Apparently, it is an eigenstate of the reflection
operator. The corresponding eigenvalue is $\sigma=1$ because
$\ket{\Omega}_{M}$ is a positive superposition of shifted basic
states \cite{LM62} and the shift operator \eqref{U} remains
invariant  with respect to the reflection. Thus, in this case the
minimal energy in the symmetric sector is lower than the
corresponding one in antisymmetric sector: $E_S=E_{S,1}<E_{S,-1}$.
As was mentioned in the Introduction, the ground state is unique
and has spin $S_\text{gs}=|S_A-S_B|$, \cite{LM62,Lieb89} i.e.,
is a singlet for even values of $N$ and triplet for odd values of
$N$. In both cases, $\sigma=1$. This is in agreement with our
results, which claim that $\sigma=(-1)^N$ for singlet ground state
and $\sigma=(-1)^{N-1}$ for triplet one. All mentioned properties
are similar to the properties of spin-1 Haldane chain because due
to ferromagnetic rung couplings, which make preferable rung
triplet states, the low-energy properties are well described  by
Haldane chain. \cite{Hida91,Wat94}

\subsection{Ferromagnetic diagonal interactions}
Consider now the ladder with ferromagnetic diagonal couplings and
antiferromagnetic rung and intrachain couplings. The low-energy
properties of this system have been investigated recently in
Ref.~\onlinecite{ZWCh00}. The model contains standard
antiferromagnetic ladder as a particular case. The frustration is
lost in this case, too. The lattice becomes bipartite of
checkerboard type and the Lieb-Mattis theorem can be applied also.
The sublattice $A$ consists of the odd sites of the
first chain and the even sites of the second chain, while the
sublattice $B$ consists of the even sites of the first chain and
the odd sites of the second one.  The spin-rotation operator
leading to a nonpositive basis is given by $U'=\exp\left(i\pi
S^z_A\right)$, where $S^z_A=\sum_l(S^{z}_{1,2l-1}+S^{z}_{2,2l})$
is the spin projection operator of $A$ sublattice. \cite{LM62} As
in the previous case, the relative ground state $\ket{\Omega}_{M}$
in $M$ subspace is expressed as a positive superposition of
the shifted basic states $
\ketcol{m_{1,1},\dots,m_{1,N}}{m_{2,1},\dots,m_{2,N}}=
U'\left(\ketcol{m_{1,1}}{m_{2,1}}\otimes\dots\otimes\ketcol{m_{1,N}}{m_{2,N}}\right)
$, where $m=\uparrow,\downarrow$ labels on-site spins. Now using
the definition of $\refl$ and $U'$ after a simple algebra, we
obtain
$$
\refl\ketcol{m_{1,1},\dots,m_{1,N}}{m_{2,1},\dots,m_{2,N}}=
e^{i\pi (S^z-2 S^z_A)}
\ketcol{m_{2,1},\dots,m_{2,N}}{m_{1,1},\dots,m_{1,N}}=
(-1)^{N-M}\ketcol{m_{2,1},\dots,m_{2,N}}{m_{1,1},\dots,m_{1,N}}.
$$
Hence, we have
$\refl\ket{\Omega}_{M}=(-1)^{N-M}\ket{\Omega}_{M}$. Thus, among
two $(M,\sigma)$ subspaces with different values of $\sigma$, the
minimal energy level in the subspace with $\sigma=(-1)^{N-M}$ is
the lowest: $E_S=E_{S,\sigma}<E_{S,-\sigma}$. The total ground
state is a spin singlet with $\sigma=(-1)^N$ reflection quantum
number in accordance with our results in \secref{sec:open}.

\subsection{Composite spin model}
The equal values of diagonal and intrachain couplings
($J_l^\times= J_l^\parallel$) correspond to so-called composite
spin model. \cite{ST86} These values of coupling parameters are
out of the range \eqref{range}, where the results of
\secref{sec:open} are valid. However, due to continuity, some
results still remain true in this limiting case. In particular,
the ordering rule is fulfilled after replacing the strict
inequality sign in \eqref{ordering} by the nonstrict one like in
the thermodynamic limit.

In contrast, the results related to the uniqueness of relative
ground states cannot be applied here. Indeed, in this case,
$J^{(a)}_l=0$ and the antisymmetrized term, which is responsible
for the exchanges between singlet and triplet rung states, is
absent in the Hamiltonian \eqref{h-U}. The spins of individual
rungs are now conserved quantities. \cite{Xian95} Therefore, all
singlets remain frozen at their sites and the Hamiltonian is
\emph{not connected} on $(M,\sigma)$ subspaces anymore. For
site-independent values of couplings, the ground state has been
obtained explicitly. \cite{Xian95} For small values of $J^\perp$,
it is reduced to the ground state of Haldane chain ($S=0$ for even
$N$, $S=1$ for odd $N$ and $\sigma=1$ always), while the large
values lead to the rung-singlet state
$\ket{0}\otimes\dots\otimes\ket{0}$ [$S=0$ and $\sigma=(-1)^N$].
This agrees with our results in \secref{sec:GS}. Both states
are eigenstates of the composite spin Hamiltonian, which makes the
level crossing at some critical point $J^\perp_c(N)<2J^\times$
inevitable. For even values of $N$, the degeneracy disappears under
a weak deviation from the composite spin point, as has been
discussed in \secref{sec:GS}. However, the two lowest levels still
approach closely, as is shown in \figref{fig:crossing20}.

\section{Ordering in frustrated ladder with boundary impurity}
\label{sec:impurity}

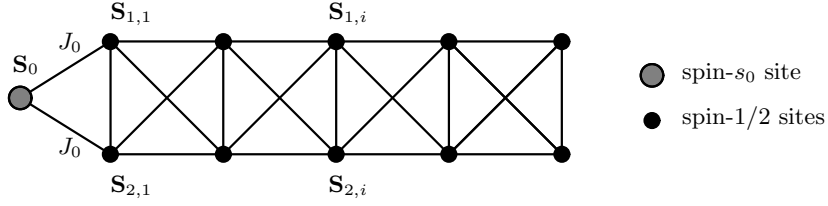
\begin{figure}
\setlength{\unitlength}{1mm}
\begingroup\makeatletter\ifx\SetFigFont\undefined%
\gdef\SetFigFont#1#2#3#4#5{%
  \reset@font\fontsize{#1}{#2pt}%
  \fontfamily{#3}\fontseries{#4}\fontshape{#5}%
  \selectfont}%
\fi\endgroup%
{\renewcommand{\dashlinestretch}{30}

\begin{picture}(110,25)(0,-5)

\allinethickness{1.000pt}%

\drawline(3,7.5)(15,0)
\drawline(3,7.5)(15,15)

\drawline(15,0)(75,0)
\drawline(15,15)(75,15)

\drawline(15,0)(15,15)
\drawline(30,0)(30,15)
\drawline(45,0)(45,15)
\drawline(60,0)(60,15)
\drawline(75,0)(75,15)

\drawline(15,0)(30,15)
\drawline(15,15)(30,0)
\drawline(60,0)(75,15)
\drawline(60,15)(75,0)
\drawline(30,0)(45,15)
\drawline(30,15)(45,0)
\drawline(45,0)(60,15)
\drawline(45,15)(60,0)
\drawline(60,0)(75,15)
\drawline(60,15)(75,0)

\texture{44555555 55aaaaaa aa555555 55aaaaaa aa555555 55aaaaaa aa555555 55aaaaaa
    aa555555 55aaaaaa aa555555 55aaaaaa aa555555 55aaaaaa aa555555 55aaaaaa
    aa555555 55aaaaaa aa555555 55aaaaaa aa555555 55aaaaaa aa555555 55aaaaaa
    aa555555 55aaaaaa aa555555 55aaaaaa aa555555 55aaaaaa aa555555 55aaaaaa }
\put(3,7.5){\shade\ellipse{3}{3}}
\put(3,7.5){\ellipse{3}{3}}
\put(15,0){\blacken\ellipse{2}{2}}
\put(15,15){\blacken\ellipse{2}{2}}
\put(30,0){\blacken\ellipse{2}{2}}
\put(30,15){\blacken\ellipse{2}{2}}
\put(45,0){\blacken\ellipse{2}{2}}
\put(45,15){\blacken\ellipse{2}{2}}
\put(60,0){\blacken\ellipse{2}{2}}
\put(60,15){\blacken\ellipse{2}{2}}
\put(75,0){\blacken\ellipse{2}{2}}
\put(75,15){\blacken\ellipse{2}{2}}

\put(2,11){$\s_0$}
\put(8,14){$J_0$}
\put(8,0){$J_0$}

\put(15,18){$\s_{1,1}$}
\put(15,-5){$\s_{2,1}$}


\put(44,18){$\s_{1,i}$}
\put(44,-5){$\s_{2,i}$}

\put(87,4.5){\blacken\ellipse{2}{2}}
\put(87,10.5){\shade\ellipse{3}{3}}
\put(87,10.5){\ellipse{3}{3}}

\put(91,4){spin-1/2 sites}
\put(91,10){spin-$s_0$ site}

\end{picture}
}
\caption{\label{fig:bounladder}
Spin-1/2 frustrated ladder with boundary impurity spin $s_0$.}
\end{figure}

Consider frustrated spin-1/2 ladder \eqref{h} with an impurity spin $s_0$
coupled antiferromagnetically to two spins of the first rung,
 as was shown on \figref{fig:bounladder}.
The original Hamiltonian \eqref{h} is supplemented by boundary
term $H_\text{imp}=J_0\,\s_0\cdot\s^{(s)}_1$, $J_0>0$, which
preserves the reflection symmetry.

\subsection{Relative ground states}
After the rotation of odd-rung spins on angle $\pi$ around $z$
axis by means of unitary operator \eqref{U}, the impurity-ladder
interaction term acquires the form.
$$
\tilde
H_\text{imp}=UH_\text{imp}U^{-1}=-\frac{J_0}2(S^{+}_0S^{(s)-}_{1}+S^{-}_0S^{(s)+}_{1})
+J_0 S^z_0S^{(s)z}_1.
$$
We will work here in a basis, which is a natural extension of the
basis \eqref{new-basis}, used before for the ladder with free
boundary conditions:
\begin{equation}
\label{boundary-basis}
\ket{m_0}\otimes\ket{m_1,\dots,m_N}=(-1)^{[N_0/2]+N_{0\tilde0}}\ket{m_0}\otimes\ket{m_1}\otimes
\ldots\otimes\ket{m_N}.
\end{equation}
Here $\ket{m_0}:=\ket{s_0,m_0}$ ($m_0=-s_0,\dots,s_0-1,s_0$) is
the usual basis of spin-$s_0$ multiplet. All nonvanishing matrix
elements of $S_0^\pm$ are positive:
$\bra{m_0}S^+_0\ket{m_0-1}=\bra{m_0-1}S^-_0\ket{m_0}=\sqrt{(s_0+m_0)(s_0-m_0+1)}>0$.
It is easy to see that the spin exchanges due to the
boundary term do not
affect on the sign factor $(-1)^{[N_0/2]+N_{0\tilde0}}$.
This together with \eqref{ssym} implies that
$\tilde H_\text{imp}$ has only nonpositive off-diagonal elements
in the basis \eqref{boundary-basis}.

 The bulk part \eqref{h-U} of
the Hamiltonian is also nonpositive in this basis as was proved
already in \secref{sec:open}. Thus, the entire Hamiltonian $\tilde
H+\tilde H_\text{imp}$ has no positive off-diagonal elements, too.

It is easy to ensure that its restriction to any ($M,\sigma$) subspace
with fixed spin projection $M$ and reflection $\sigma$ quantum numbers
is connected.
This can be shown using the arguments similar to those applied
before for the ladder without impurity spin.

Applying again Perron-Frobenius theorem, we come to the conclusion that
the relative ground state of $\tilde H+\tilde H_\text{imp}$ of each ($M,\sigma$) subspace
is unique and is a positive superposition of \emph{all} basic states \eqref{boundary-basis}:
\begin{equation}
\label{boundary-gs}
\ket{\Omega}_{M,\sigma}=\sum_{\substack{\sum_lm_l=M\\
(-1)^{N_0}=\sigma}} \omega_{m_0\dots
m_N}\ket{m_0}\otimes\ket{m_1,\ldots,m_{N}}, \qquad
\omega_{m_0\dots m_N}>0.
\end{equation}

Now turn to the determination of the spin $S_{M,\sigma}$ of this
state. As before, in order to do this, we will construct a state
with certain value of spin and positive or vanishing coefficients
in its decomposition over the basic states \eqref{boundary-basis}.
In this case, it will overlap with the relative ground state
\eqref{boundary-gs}. Then the uniqueness of the relative ground
state would indicate that both states have the same spin value.

Due to the spin reflection symmetry, both states $\ket{\Omega}_{\pm
M,\sigma}$ have the same spin quantum number. Thus, without any
restriction, we can consider $M\ge0$ values only. First, we consider
the values $M\ge s_0$ and
 will prove below that  $S_{M,\sigma}=M$.

In $\sigma=(-1)^{N-M+s_0}$ sector, there is a simple highest state
$\ket{s_0}\otimes\kett{\underbrace{1,\dots,1}_{M-s_0},0,\dots,0}$
of a multiplet with spin $S=M$. It coincides up to an unessential
sign factor with one of the basic states \eqref{boundary-basis}.
Therefore, $S_{M,\sigma}=M$ for this case.

For the opposite values of the reflection quantum number
[$\sigma=(-1)^{N-M+s_0-1}$], we take as an indicator the state
$$
\ket{\psi'}=\left(\sqrt{\frac{s_0}{s_0+1}}\,\ket{s_0}\otimes\ket{\tilde0}
+\frac{1}{\sqrt{s_0+1}}\,\ket{s_0-1}\otimes\ket{1}\right) \otimes
\kett{\underbrace{1,\dots,1}_{M-s_0},0,\dots,0},
$$
which can be considered as a generalization of the state
$\ket{\psi}$ used before for bulk ladder. Again, $\ket{\psi'}$ is
a \emph{positive} superposition of two basic states from
\eqref{boundary-basis}. As has been done before for state
$\ket{\psi}$, its spin can be detected going back to the original
$H$ representation: $\ket{\psi'}\to U^{-1}\ket{\psi'}$. As a
result, just the sign of the second term in the brackets is
changed. It is easy to check that $ \sqrt{\frac{s_0}{s_0+1}}\,
\ket{s_0}\otimes\ket{\tilde0} -\sqrt{\frac{1}{s_0+1}}\,
\ket{s_0-1}\otimes\ket{1}$ is the highest state of spin-$s_0$
multiplet, which appears in the decomposition of the tensor
product of two multiplets with spins $s_0$ and $1$. Thus, the spin
of the entire shifted state $U^{-1}\ket{\psi'}$ is again  $M$.

Next, we consider the values $0\le M<s_0$, which are essential for
$s_0\ge1$. We will show that in this case the relative ground
state in $(M,\sigma)$ subspace belongs to a spin
$S_{M,\sigma}=s_0$ multiplet if $\sigma=(-1)^{N}$; otherwise it
belongs to a spin $S_{M,\sigma}=s_0-1$ multiplet. The first case
is easy to prove using for the test the basic state
$\ket{M}\otimes\ket{0,\dots,0}$ of spin $s_0$. For the second case
we choose as a test the state
$\ket{\psi''}=\ket{\chi}\otimes\underbrace{\ket{0}\otimes\ldots\otimes\ket{0}}_{N-1}$,
where
\begin{equation*}
\begin{split}
\ket{\chi}=&\sqrt{\frac{(s_0-M+1)(s_0-M)}{2s_0(2s_0+1)}}\,\ket{M-1}\otimes\ket{1}
+\sqrt{\frac{(s_0-M)(s_0+M)}{s_0(2s_0+1)}}\,\ket{M}\otimes\ket{\tilde0}
\\
+&\sqrt{\frac{(s_0+M+1)(s_0+M)}{2s_0(2s_0+1)}}\,\ket{M+1}\otimes\ket{-1}.
\end{split}
\end{equation*}
It is easy to see that  $\ket{\psi''}$ is a positive superposition
of three states from the basic set \eqref{boundary-basis}.
The action of $U^{-1}$ on $\ket{\psi''}$ changes only the sign of
the first and last terms of $\ket{\chi}$ giving rise to a state of
spin $S=s_0-1$ multiplet, as  is easy to verify by applying
 $(\s_0+\s^{(s)}_1)^2$ to it.

Finally, summarizing the results obtained above, we arrive at the
following conclusion.
\emph{
\begin{itemize}
\item For the ladder with diagonal interactions obeying
\eqref{range} and antiferromagnetically coupled boundary impurity
$s_0$ the relative ground state in $(M,\sigma)$ subspace is
nondegenerate and belongs to a multiplet with the spin
\begin{equation}
\label{Srelgsb}
S_{M,\sigma}=
\begin{cases}
|M| & \text{if $|M|\ge s_0$}\\
s_0 & \text{if $|M|<s_0$ and $\sigma=(-1)^N$}\\
s_0-1 & \text{if $|M|<s_0$ and $\sigma=(-1)^{N-1}$.}
\end{cases}
\end{equation}
\end{itemize}
}
Note that for $s_0=1/2$ value of impurity spin, only the first
line holds and we have simply $S_{M,\sigma}=|M|$.

\subsection{Ordering rule and ground state}
Using the spin value  of the relative ground state \eqref{Srelgsb} and its uniqueness
proved above, one can compare the minimal energy levels $E_{S,\sigma}$ of the sectors with fixed
spin and reflection quantum numbers and study their degeneracy
just in the same way as was done above for the open ladder.

For $\sigma=(-1)^N$, it is clear from \eqref{Srelgsb} that  the energy
level $E_{S,\sigma}$ is nondegenerate for any $S\ge s_0$.
The uniqueness of the relative ground state suggests that all
levels with spin $S>S_{M,\sigma}$ are above the level with
$S=S_{M,\sigma}$, which  contains the state $\ket{\Omega}_{M,\sigma}$ itself.
Thus, $E_{S,\sigma}$ is a monotone increasing function of spin in the
range $S\ge s_0$. For $S<s_0$, the level degeneracy and ordering are not
clear. All relative ground states $\ket{\Omega}_{M,\sigma}$ with $|M|\le s_0$
constitute a spin-$s_0$ multiplet having the lowest-energy value among all states
with the reflection parity $\sigma=(-1)^N$.

For $\sigma=(-1)^{N-1}$, the level $E_{S,\sigma}$ is nondegenerate for any $S\ge s_0-1$.
The lowest energy $E_{S,\sigma}$ is a monotone increasing function of spin in the
range $S\ge s_0-1$. In the case of $s_0\ge1$, all relative ground states
$\ket{\Omega}_{M,\sigma}$ with $|M|\le s_0-1$
are combined into $S=s_0-1$ multiplet having the minimal energy among the states with
$\sigma=(-1)^{N-1}$. For $s_0=1/2$, the lowest level increases with spin everywhere.
The following statement sums up the discussions above.
\emph{
\begin{itemize}
\item
For the ladder with diagonal interactions obeying $J^\parallel_l>|J^\times_l|$
and antiferromagnetically coupled
boundary impurity $s_0$ the lowest energy levels in sectors with fixed value of spin $S$ and
reflection $\sigma$ quantum numbers are ordered according to the rule:
\begin{equation}
\label{boundary-ordering}
E_{S_1,\sigma}>E_{S_2,\sigma}:
\quad
\begin{cases}
\text{for $\sigma=(-1)^N$}       & \text{if either $S_1>S_2\ge s_0$  or  $S_1<S_2=s_0$}\\
\text{for $\sigma=(-1)^{N-1}$}   & \text{if either $S_1>S_2\ge s_0-1$  or  $S_1<S_2=s_0-1$.}
\end{cases}
\end{equation}
\item
The ground state of the model in entire $\sigma=(-1)^N$ sector is a
 $S=s_0$ multiplet, while in  entire $\sigma=(-1)^{N-1}$ sector is a
 $S=|s_0-1|$ multiplet. In both cases, it is unique.
\end{itemize}
}
The absolute value sign is used in order to take into consideration
the impurity with $s_0=1/2$ also. In that case, the ordering rule \eqref{boundary-ordering}
is simplified  because in that case $S=1/2$ is the lowest value of the total spin and the
relation $E_{S_1,\sigma}>E_{S_2,\sigma}$ holds for any $\sigma$ and any two spins obeying
$S_1>S_2$.
The lowest-energy states in both symmetric and antisymmetric sectors are unique
spin doublets.

As for the ladder model without impurity, one cannot compare energy
levels related to different reflection quantum numbers.
Nevertheless, one can gain information about the total ground
state of the model from the ordering rule established above. The
total ground state is nondegenerate and either belongs to
$\sigma=(-1)^N$ sector and, hence, has spin value $S=s_0$ or
belongs to $\sigma=(-1)^{N-1}$ and has spin $S=|s_0-1|$. If the
lowest energy levels in both sectors coincide, which can occur,
for example, due to some additional symmetry presented in the
model, then the ground state becomes degenerate and is the
superposition of two multiplets with spin values $s_0$ and
$|s_0-1|$.
Therefore,  the ground state could be \emph{at most}
doubly degenerate.

In contrast to the model with free boundary conditions studied in \secref{sec:open},
here we cannot apply the Lieb-Schupp results \cite{LS99,LS00}
in order to choose the valid quantum numbers for the ground state. The reason is that their
approach cannot be used for a reflection-symmetric spin system if some spins
are positioned on the symmetry axis.

\subsection{An example: Periodic spin-1/2 chain with odd number of spins}
The translationally invariant periodic spin-1/2 chain with
\emph{odd} $N'=2N+1$ number of spins is a particular case of the general class
of frustrated ladders with boundary spin, considered in this section.
It corresponds to $s_0=1/2$ and
$J_0=J^\parallel_l=J^\perp_N$ with vanishing values for other coupling coefficients.
The model is integrable by Bethe ansatz \cite{Kah95} like
its more familiar even-site counterpart, but many of its properties
differ from those of even-site chain. \cite{Schmidt03} In
particular, the even-site chain is bipartite and Lieb-Mattis
theorem is valid in this case, while the odd-site chain is a
frustrated system. Recently, the classical ground state of the last
model has been constructed and the lowest energy has been obtained
exactly. \cite{SchL03}

The system possesses an additional symmetry: it remains invariant with respect to
the cyclic translation by one site $T$.
The translation and reflection operators satisfy the commutation relation $\refl T=T^{-1}\refl$.
Thus, if $\ket{\psi}$ is an eigenstate of $T$ with eigenvalue $e^{i\phi}$
then the reflected state $\refl\ket{\psi}$ is also an eigenstate
with eigenvalue $e^{-i\phi}$. Hence, for all values of momentum except  $\phi=0,\pi$,
the energy levels of periodic chain are at least twofold degenerate.

The relative ground states of even-site translationally invariant
chain have just the exceptional values of momenta ($0$ or $\pi$),
which is in agreement with their uniqueness. In contrast, for
odd-site chain, the exact solution shows that relative ground state
in $S^z=M$ subspace is \emph{exactly} doubly degenerate with two
opposite momenta $\phi=\pm \pi(M+1/N')$. \cite{Kah95} This is in
agreement with our results, which assert that it should be
\emph{at most} doubly degenerate. Therefore, the relative ground
states in symmetric and antisymmetric $M$ subspaces have the same
energy levels. The reason of the degeneracy is the translation
symmetry, which mixes these two states. Hence, the lowest levels
in symmetric and antisymmetric sectors coincide
($E_{S,1}=E_{S,-1}=E_S$) except, of course, $M=\pm N'/2$
subspaces, where only one ferromagnetic state presents. Moreover,
antiferromagnetic ordering rule ($E_{S_1}>E_{S_2}$ if $S_1>S_2$)
is fulfilled like for even-site chains. The total ground state is
the combination of two spin doublets with different reflection
quantum numbers.

\section{Periodic ladder}
\label{sec:periodic}

Consider now the frustrated ladder with periodic boundary
conditions. The bulk Hamiltonian \eqref{h} is supplemented by the
term
$$
H_\text{per}=
J^\parallel_N(\s_{1, N}\cdot\s_{1, 1}+\s_{2, N}\cdot\s_{2, 1})
+
J^\times_N(\s_{1,N}\cdot\s_{2,1}+\s_{1,1}\cdot\s_{2,N}),
$$
which describes the interactions between the spins of the first
and last rungs. Following \eqref{range} we set the condition
$J^\perp_N>|J^\times_N|$ on the boundary coupling constants. In the
expression of the Hamiltonian, in terms of symmetrized and
antisymmetrized spin operators \eqref{hsa}, the boundary term
contributes as $H_\text{per}=J^s_N\,
\s^{(s)}_N\cdot\s^{(s)}_{1}+J^a_N\, \s^{(a)}_N\cdot\s^{(a)}_{1}$,
where the coefficients $J^s_N$, $J^a_N$ are given in \eqref{Jsa}.
Both are antiferromagnetic.

Under the action of unitary shift operator \eqref{U}, the total
Hamiltonian acquires the form $\tilde H+\tilde H_\text{per}$,
where $\tilde H$ is derived already in \eqref{h-U} and
\begin{equation}
\label{h-spm-b}
\begin{aligned}
\tilde
H_\text{per}=UH_\text{per}U^{-1}&=\frac{(-1)^{N-1}}{2}\left(J^s_N
S^{(s)+}_1 S^{(s)-}_{N}+J^s_N S^{(s)-}_1 S^{(s)+}_N +J^a_N
S^{(a)+}_N S^{(a)-}_{1}+ J^a_N S^{(a)-}_N S^{(a)+}_{1}\right)
\\
&+J^s_N S^{(s)z}_1 S^{(s)z}_N+J^a_N S^{(a)z}_1 S^{(a)z}_N.
\end{aligned}
\end{equation}

The boundary term $\tilde H_\text{per}$  produces positive off-diagonal
elements in the basis \eqref{new-basis}, in which the bulk part
$\tilde H$ of the total Hamiltonian is negative.
One reason is the sign factor $(-1)^{N-1}$ in
\eqref{h-spm-b} due to which positive off-diagonal elements appear
for even values of $N$. Another more important reason is that the
sign factor $(-1)^{N_{0\tilde0}}$ in front of basic states
\eqref{new-basis} is essentially nonlocal. Remember that
$N_{0\tilde0}$ counts the number of such $(\ket{0},\ket{\tilde0})$
pairs that $\ket{0}$ is positioned on the left-hand side from
$\ket{\tilde0}$. The exchange between the first and last rung
spins  can produce an uncertain sign
factor, which depends on intermediate rung states. Below, we will
derive the conditions, under which the basis \eqref{new-basis}
nevertheless remains nonpositive.

The term $S^{(s)z}_1 S^{(s)z}_N$ is diagonal in the basis
\eqref{new-basis}. The other terms produce off-diagonal elements.
First, consider  the  matrix elements, which are generated by the terms with $S^{(a)z}$.
They do not depend on the parity of $N$. Using \eqref{sz} and \eqref{new-basis}
it is easy to check that
\begin{equation}
\label{sz-per}
\bra{\tilde0,\dots,\tilde0}S^{(a)z}_1S^{(a)z}_N\ket{0,\dots,0}=
\bra{0,\dots,\tilde0}S^{(a)z}_1S^{(a)z}_N\ket{\tilde0,\dots,0}=(-1)^{N_0+N_{\tilde0}+1},
\end{equation}
where the dots indicate all intermediate rung states, which have to be the same for the
bra and ket states to ensure that the matrix element is nonzero. Here, $N_0$ and $N_{\tilde0}$
are the numbers of rung-singlet and $S^z=0$ triplet states correspondingly.

Next, we consider the matrix elements generated by lowering-rising spin operators in \eqref{h-spm-b}.
Taking into account \eqref{ssym}, \eqref{spm}, and \eqref{new-basis}, we obtain
\begin{equation}
\begin{aligned}
\label{spm-per}
\bra{\tilde0,\dots,\tilde0}S^{(s)\mp}_1S^{(s)\pm}_N\ket{\pm1,\dots,\mp1}=
\bra{\pm1,\dots,\tilde0}S^{(s)\pm}_1S^{(s)\mp}_N\ket{\tilde0,\dots,\pm1}=(-1)^{N_0}\cdot2,
\\
\bra{0,\dots,0}S^{(a)\mp}_1S^{(a)\pm}_N\ket{\pm1,\dots,\mp1}=
\bra{\pm1,\dots,0}S^{(a)\pm}_1S^{(a)\mp}_N\ket{0,\dots,\pm1}=(-1)^{N_{\tilde0}}\cdot2.
\end{aligned}
\end{equation}
The matrix elements \eqref{sz-per} and \eqref{spm-per} together with
their transpositions are the only off-diagonal elements generated by $\tilde H_\text{per}$.
They will be negative if $(-1)^{N_{\tilde0}+N_0}=(-1)^{N+N_0}=(-1)^{N+N_{\tilde0}}=1$
because the values of couplings $J^s_N$ and $J^a_N$ are positive.
Using the relations $\sigma=(-1)^{N_0}$ for reflection quantum number $\sigma$
and $(-1)^M=(-1)^{N-N_{\tilde0}-N_0}$ for the spin projection quantum number $M$,
we can rewrite the last equations  as
\begin{equation}
\label{sigma-N-M}
\sigma=(-1)^N=(-1)^M.
\end{equation}
Remember now that the bulk part of the Hamiltonian has only
nonpositive off-diagonal elements in the basis \eqref{new-basis}
for any $M$ and $\sigma$ (see \secref{sec:open}). Therefore, the
entire Hamiltonian is nonpositive only on those $(M,\sigma)$
subspaces (where $S^z=M$ and $\refl=\sigma$), which are subjected
to the condition \eqref{sigma-N-M}. Moreover, in any such subspace,
its matrix is connected because $\tilde H$ is connected as was
already proven. Thus, according to Perron-Frobenius theorem, the
relative ground state in each $(M,\sigma)$ subspace with quantum
numbers obeying \eqref{sigma-N-M} is nondegenerate and is a
positive superposition of all basic states \eqref{gs}. The last
fact has been used in \secref{sec:open} in order to determine the
spin $S_{M,\sigma}$ of the relative ground state for the open
ladder. So, here, the formula \eqref{Srelgs} can be applied too,
and, actually, it is reduced just to $S_{M,\sigma}=|M|$ since the
equations \eqref{sigma-N-M} exclude the  exceptional case of $M=0$
and $\sigma=(-1)^{N-1}$. The outcome is as follows.
\emph{
\begin{itemize}
\item
For the periodic frustrated ladder with even  number of rungs, the relative ground state
in any ($M=\emph{even},\sigma=1$) subspace is unique and belongs to a spin-$|M|$ multiplet.
For the ladder with odd number of rungs,
the same is true  in any ($M=\emph{odd},\sigma=-1$) subspace.
\end{itemize}
} As a consequence, the lowest energy levels $E_{S,\sigma}$ among
all states with fixed spin $S$ and reflection $\sigma$ quantum
numbers are nondegenerate for $\sigma=1$ and even values of $S$ as
well as for  $\sigma=-1$ and odd values of $S$. Moreover, we get a
partial antiferromagnetic ordering of the lowest energy levels
$E_{S,\sigma}$.
\emph{
\begin{itemize}
\item
For the periodic frustrated ladder with even number of rungs, the relation $E_{S_1,1}>E_{S_2,1}$
holds for any \emph{even} spin $S_2$ and any spin $S_1$ such that $S_1>S_2$.
The ground state in entire symmetric ($\sigma=1$) sector is a unique spin singlet.
\item
For ladder with odd number of rungs, $E_{S_1,-1}>E_{S_2,-1}$ for any \emph{odd} spin
$S_2$ and any spin $S_1$ satisfying $S_1>S_2$.
\end{itemize}
}
For odd  $N$, the lowest state in the antisymmetric may be a unique spin triplet,
a spin singlet(s), or a singlet(s)$\oplus$\,triplet representation.

Remember that in this paper we consider the ladders with
coupling obeying $J^\parallel_l>|J^\times_l|$. The ladder with
periodic boundary term under certain supplementary condition on
coupling constants becomes a part of more general class of
reflection-symmetric models investigated recently by Lieb and
Schupp. \cite{LS99,LS00} This fact has been used in
\secref{sec:open} in the case of free boundaries and is discussed in
detail in Appendix. For the coupling values obeying $J^\perp_l>
|J^\times_{l-1}|+|J^\times_l|$, all ground states become
spin singlets, and among them there is one with $\sigma=(-1)^N$.
Moreover, for antiferromagnetic diagonal interactions (i.e., if all
$J^\times_l>0$) the nonstrict inequality sign can be used in the
last inequality instead of strict one [see \eqref{cond-all} in
Appendix]. In particular, the translationally invariant frustrated
ladder with the antiferromagnetic couplings obeying $J^\perp\ge
2J^\times$ and $J^\parallel>J^\times$  has only spin-singlet
ground states as well as is subjected to the ordering rule proved
in this section.

For the ladders with even rungs, the aforementioned properties are in good agreement with
our results obtained in this section.
For the ladders with odd rungs, they even strengthen the partial ordering rule.
The relation $E_{S_1,-1}>E_{S_2,-1}$, holding for any odd $S_2$ and any $S_1$ such that
$S_1>S_2$, becomes valid for $S_2=0$, too.
The question of the  degeneracy of the level $E_{0,-1}$ still remains open.

Unlike in open ladder case, here
we did not obtain  definite exact results related
to the degeneracy degree and spin quantum number of the total ground state.
The main reason is the absence of exact results related to $\sigma=(-1)^{N-1}$
sector in the case when the periodic boundary term presents.

\section{Summary and conclusion}

We have generalized Lieb-Mattis energy level ordering rule for spin systems
on bipartite lattices \cite{LM62} to the frustrated spin-1/2 ladder model.
The model consists of two coupled antiferromagnetic chains frustrated by diagonal interactions
and possesses reflection symmetry with respect to the longitudinal axis.
The spin exchange coupling constants depend on site and are subjected to the relations
\eqref{range}.
We have considered the finite-size system with free boundaries, the system with any impurity
spin  attached to one boundary as well as the ladder with periodic boundary conditions.
Below, we describe briefly the results obtained in this paper.

The total spin $S$ and reflection parity $\sigma=\pm1$ are good
quantum numbers. So, the Hamiltonian remains invariant on
individual sectors with fixed values of both quantum numbers. For
the open ladder and ladder with impurity, we have established that
the lowest-energy levels $E_{S,\sigma}$ of these sectors with the
same value of $\sigma$ are ordered antiferromagnetically. More
precisely, the relation $E_{S_1,\sigma}>E_{S_2,\sigma}$ holds for
any two spins satisfying $S_1>S_2\ge S_\text{gs}(\sigma)$. Here,
$S_\text{gs}(\sigma)$ is the spin value of the \emph{unique} lowest level multiplet
among all multiplets with the reflection parity $\sigma$:
$$
S_\text{gs}(\sigma)=
\begin{cases}
s_0 & \text{if $\sigma=(-1)^N$}\\
|s_0-1| & \text{if $\sigma=(-1)^{N-1}$},
\end{cases}
$$
where $N$ is the number of rungs and $s_0$ is the impurity spin
value. The case $s_0=0$ corresponds to the open ladder without
impurity. We have also proven that all levels
$E_{S,\sigma}$ with $S\ge S_\text{gs}(\sigma)$ are nondegenerate,
which means that only one multiplet  exists on the corresponding level.
Note that  $S_\text{gs}(\sigma)$ coincides for $\sigma=1$ with the ground-state
spin $S_\text{gs}$ of the Haldain chain obtained after replacing each rung with the
spin $s=1$. The last model is bipartite and the Lieb-Mattis formula
$S_\text{gs}=|S_A-S_B|$, mentioned in the Introduction, gives just the value
of $S_\text{gs}(1)$.

 In contrast to bipartite
spin systems, the lowest levels in the symmetric ($\sigma=1$) and
antisymmetric ($\sigma=-1$) sectors cannot be compared with each
other, at least by means of our approach. Therefore, in general,
the total ground state of the model is a unique multiplet with
spin $s_0$ or $|s_0-1|$. If the lowest levels in both sectors
coincide, then the degeneracy occurs and the ground state is the
superposition of both multiplets. In this regard, our results claim that the
ground state can be \emph{at most} doubly degenerate.

For more restrictive values of couplings given in Appendix, the
frustrated ladder with free boundaries ($s_0=0$) fits the class of
reflection-symmetric spin systems, all ground states of which are
spin singlets. \cite{LS99,LS00} Combining with our results, this
implies that the ground state is a \emph{unique} spin singlet in
this case. This property is true, in particular, for frustrated
ladder with site-independent antiferromagnetic couplings obeying
$J^\perp\ge 2J^\times$ and $J^\parallel>J^\times$.

For ladder with periodic boundary conditions, we have proven a
weaker ordering rule. Namely, for the ladder with even (odd)
number of rungs the ordering $E_{S_1,\sigma}>E_{S_2,\sigma}$ if
$S_1>S_2$ is established exactly for $\sigma=1$ ($\sigma=-1$) and
even (odd) values of spin $S_2$ only. The degeneracy of the ground
state and its total spin value remain  open questions in this
case.

\begin{acknowledgments}
The author expresses his gratitude to A.~A.~Nersesyan for very
useful discussions. He is thankful also to the Abdul Salam
International Center for Theoretical Physics (ICTP), where this
work was started, for hospitality.

The work was supported by the Volkswagen Foundation of Germany,
grants INTAS-03-51-4000 and INTAS-05-7928.

\end{acknowledgments}

\appendix
\section{Application of Lieb-Schupp approach to frustrated spin ladder}

Recently, Lieb and Schupp \cite{LS00} proved exactly that a
reflection-symmetric spin system with antiferromagnetic crossing
bonds has at least one spin-singlet ground state.
Moreover, under certain additional conditions, all ground states
become singlets.
Frustrated spin-1/2 ladder \eqref{h} is an example of reflection-symmetric
system (see \figref{fig:ladder}). The Hamiltonian in the
Lieb-Schupp form is
\begin{equation}
\label{h-LS} H=\sum_{l=1}^N \gamma_l^2 \s_{1,l}\cdot\s_{2,l}+
\sum_{l=1}^N
(\alpha_l\s_{1,l}+\beta_l\s_{1,l+1})\cdot(\alpha_l\s_{2,l}+\beta_l\s_{2,l+1})
+\sum_{l=1}^{N} J_l (\s_{1, l}\cdot\s_{1,l+1}+ \s_{2,
l}\cdot\s_{2,l+1}).
\end{equation}
Here,  $\alpha_l,\beta_l,\gamma_l,J_l$ are arbitrary real coefficients.
The boundary conditions are
$\alpha_N=\beta_N=J_N=0$ for open ladder and $\s_{\delta,N+1}=\s_{\delta,1}$
($\delta=1,2$)
for periodic ladder. There are two kinds of
bond in \eqref{h-LS}: the first sum consists of rung bonds while
the second sum contains square bonds. Comparing \eqref{h-LS} with
\eqref{h} one can express the spin coupling constants in \eqref{h}
 in terms of Lieb-Schupp
coefficients:
\begin{equation}
\label{J-alpha} J^\parallel_l=J_l, \qquad
J^\perp_l=\alpha_{l}^2+\beta_{l-1}^2+\gamma_l^2, \qquad
J^\times_l=\alpha_l\beta_l.
\end{equation}
In the second equation above the boundary conditions
$\beta_{0}=0$ and $\beta_0=\beta_N$ are used for open and periodic ladders
correspondingly.
Note that the value of interchain rung coupling is always antiferromagnetic.

Exploiting the reflection symmetry, Lieb and Schupp proved that
there is a ground state of \eqref{h-LS}, which overlaps with rung-singlet
state $\ket{0,\dots,0}$ (Ref.~\onlinecite{LS99}) and, more generally,
with the canonical singlet state. \cite{LS00} This means that the
spin ladder Hamiltonian with diagonal interactions \eqref{h}
possesses a spin-singlet ground state with $\sigma=(-1)^N$
reflection symmetry for any value of $J^\parallel_l$ provided that
the remaining two couplings are subjected to the relation
\begin{equation}
\label{range-LS} J^\perp_l\ge\alpha_{l}^2+\beta_{l-1}^2, \qquad
J^\times_l=\alpha_l\beta_l
\end{equation}
for some $\alpha_l,\beta_l$. Setting several restrictions on
auxiliary parameters $\alpha_l,\beta_l,\gamma_l$, one can simplify
\eqref{J-alpha} and \eqref{range-LS}  significantly and get the
constraints containing interaction couplings only. Of course,
those restrictions narrow the total region  in phase space, where
Lieb-Schupp result is valid.

Consider the simplest and most familiar case of ladder with
site-independent couplings. We  set $\alpha_l=\alpha$,
$\beta_l=\beta$, $\gamma_l=0$ for periodic ladder, where
$\alpha,\beta$ are some real parameters. For open ladder, we choose
the boundary values of $\gamma_l$ parameter as
$\gamma_1=\beta,\gamma_N=\alpha$. It is easy to verify that both
for open and periodic boundary conditions, all values of interchain
couplings satisfying $J^\perp=\alpha^2+\beta^2$,
$J^\times=\alpha\beta$ lie inside the general region in phase
space defined by \eqref{range-LS}. Hence, there is a
$\sigma=(-1)^N$ singlet among the ground states if two interchain
coupling constants satisfy
\begin{equation}
\label{cond-trans}
J^\perp\ge 2|J^\times|.
\end{equation}
This relation can be generalized to the ladder model with
site-dependent couplings. Setting $\alpha_l=\pm\beta_l$, one
reduces the general region \eqref{range-LS} to
\begin{equation}
\label{cond}
 J^\perp_l\ge  |J^\times_{l-1}|+|J^\times_l|.
\end{equation}
Here, as before, the boundary conditions $J^\times_{0}=J^\times_N=0$
and $J^\times_{0}=J^\times_N$ are applied
for open and periodic ladders correspondingly.

Another important property of reflection-symmetric spin systems
established exactly in Refs.~\onlinecite{LS99} and
\onlinecite{LS00} is that for any crossing bond, the expectation
of the spin of its all sites, weighted by their coefficients,
vanishes for \emph{any} ground state $\ket{\Omega}$. Using this
feature of ground states, it was proved that in a system with
sufficient symmetry, so that every spin can be considered to be
involved in a crossing bond, \emph{all} ground states are
spin singlets.

As was already mentioned, for frustrated ladder, there are rung
bonds and square bonds, which are described correspondingly by the
first and second term in \eqref{h-LS}. The aforementioned
condition means that $\bra{\Omega}\s^{(s)}_l\ket{\Omega}=0$ if
$\gamma_l\ne 0$ and
$\bra{\Omega}\alpha_l\s^{(s)}_l+\beta_l\s^{(s)}_{l+1}\ket{\Omega}=0$,
where $\s^{(s)}_l$ is the spin operator of $l$th rung
\eqref{S-sa}. Therefore, if $\alpha_l=\beta_l\ne0$, then the
expectation of total spin of each square bond vanishes also, i.e.,
$\bra{\Omega}\s^{(s)}_l+\s^{(s)}_{l+1}\ket{\Omega}=0$.
Applying this to the ladder with site-independent spin exchange, one can
ensure that all its ground states are singlets provided that the interchain
couplings satisfy
\begin{equation}
\label{cond-all-trans}
J^\perp\ge 2 J^\times>-J^\perp.
\end{equation}
Indeed, using aforementioned
parametrization of $J^\perp,J^\times$ and taking sum over all square
bonds, we get $(\alpha+\beta)\bra{\Omega}\s\ket{\Omega}=0$ for periodic ladder
and any ground state $\ket{\Omega}$.
For the open ladder, one must add two boundary rung bonds with coefficients
$\beta$ and $\alpha$ in order to obtain the last equation.
So, if $\alpha\ne-\beta$, then any ground state is a singlet.

Consider now the ladder with site-dependent spin exchange and set
again $\alpha_l=\pm\beta_l$ in \eqref{J-alpha}. Note that if all
$\gamma_l\ne0$, which makes strict the inequality sign in
\eqref{cond}, then due to the discussions above the mean value of
the spin of every rung vanishes for any ground state. Therefore,
all ground states are singlets if $ J^\perp_l>
|J^\times_{l-1}|+|J^\times_l|$. In fact, the inequality is
nonstrict for positive values of diagonal couplings. This becomes
apparent if we perform the sum over square bonds instead of rung
bonds using the parametrization $\alpha_l=\beta_l$. Then the
ground-state mean value of the total spin of each such bond
vanishes as has been mentioned above. Taking the sum over all
square bonds for the periodic ladder and the sum over half of
square bonds in checkerboard order for the open ladder with even
number of rungs, we obtain the zero expectation for the total spin
of the model provided that
\begin{equation}
\label{cond-all}
J^\perp_l\ge  J^\times_l + J^\times_{l-1},
\quad
J^\times_l>0.
\end{equation}

For the open ladder with odd number of rungs, one of the boundary
rung spins must be added to that sum. The ground state then is a
singlet if the corresponding coefficient [$\gamma_1$ or $\gamma_N$
in \eqref{h-LS}] is nonzero. This means that the inequality in one
of the two boundary relations in \eqref{cond} must be strict, i.e.,
$J^\perp_1> J^\times_1$ or $J^\perp_N>  J^\times_{N-1}$.

\end{document}